\documentclass[rnote]{aa}
\usepackage{graphicx}
\usepackage{txfonts}
\usepackage{natbib}
\bibpunct{(}{)}{;}{a}{}{,} 

\begin{document}

\authorrunning{Schr\"oder et al.}

\titlerunning{Quiet-Sun chromosphere as a universal phenomenon}

\title{Basal chromospheric flux and Maunder Minimum-type stars: \\
The quiet-Sun chromosphere as a universal phenomenon} 

\author
{K.-P. Schr\"oder\inst{1}\and
M. Mittag\inst{2}\and
M. I. P{\'e}rez Mart{\'i}nez\inst{1}\and
M. Cuntz\inst{3}\and
J. H. M. M. Schmitt\inst{2}   }

\institute{Departamento de Astronom\'ia, Universidad de Guanajuato,
Apartado Postal 144,, 36000 Guanajuato, Mexico,
\email{kps@astro.ugto.mx}
\and
Hamburger Sternwarte, Universit\"at Hamburg, Gojenbergsweg 112,
21029 Hamburg, Germany
\and
Department of Physics, Science Hall, University of Texas at Arlington,
Arlington, TX 76019-0059, USA  }

\date{Received \dots; accepted \dots}

\abstract{}
{We demonstrate the universal character of the quiet-Sun chromosphere
among inactive stars (solar-type and giants).  
By assessing the main physical 
processes, we shed new light on some common observational phenomena.}
{We discuss measurements of the solar Mt. Wilson S-index, obtained by the 
Hamburg Robotic Telescope around the extreme minimum year 2009, and 
compare the established chromospheric basal Ca~II~K line flux
to the Mt. Wilson S-index data of inactive (``flat activity") stars,
including giants.} 
{During the unusually deep and extended activity minimum of 2009, the Sun 
reached S-index values considerably lower than in any of its previously 
observed minima. In several brief periods, the Sun coincided exactly
with the S-indices of inactive (``flat", presumed Maunder Minimum-type) 
solar analogues of the Mt. Wilson sample; at the same time, the solar
visible surface was also free of any plages or remaining weak activity
regions.  The corresponding minimum Ca~II~K flux of the quiet Sun 
and of the presumed Maunder Minimum-type stars in the Mt. Wilson sample 
are found to be identical to the corresponding Ca~II~K chromospheric basal
flux limit.}
{We conclude that the quiet-Sun chromosphere is a universal phenomenon 
among inactive stars.  Its mixed-polarity magnetic field, generated by
a local, ``fast" turbulent dynamo finally provides a natural explanation 
for the minimal soft X-ray emission observed for inactive stars.  Given
such a local dynamo also works for giant chromospheres, albeit on larger 
length scales, i.e., $l \propto R/g$, with $R$ and $g$ as stellar radius and
surface gravity, respectively, the existence of giant spicular phenomena and
the guidance of mechanical energy toward the acceleration zone of cool 
stellar winds along flux-tubes have now become traceable. }

\maketitle

\section{Introduction}

The chromospheric and coronal heating problem is one of the
unsolved questions in the realm of solar and stellar physics.
The underlying dominant physical processes have not been
identified in detail, and we are far from having fully understood
the quite diverse observational evidence pertaining to the observed
plethora of chromospheric and coronal emissions of cool stars.
The recent lull of solar activity has raised the question to what
extent we can regard the quiet-Sun and its chromosphere as
representative for other inactive or low-activity stars, particularly
the so-called Maunder Minimum (``flat activity") Mt. Wilson stars (Baliunas 
et al. 1995), or the many inactive giants with emission line fluxes
close to the basal chromospheric flux limit as discussed by, e.g.,
P{\'e}rez Mart{\'i}nez et al. (2011).  Can we even obtain 
insight into the guiding mechanisms of extended chromospheres and cool winds 
of giants by analyzing the quiet Sun?

\begin{figure*}  
\centering
\resizebox{0.9\hsize}{!}{\includegraphics{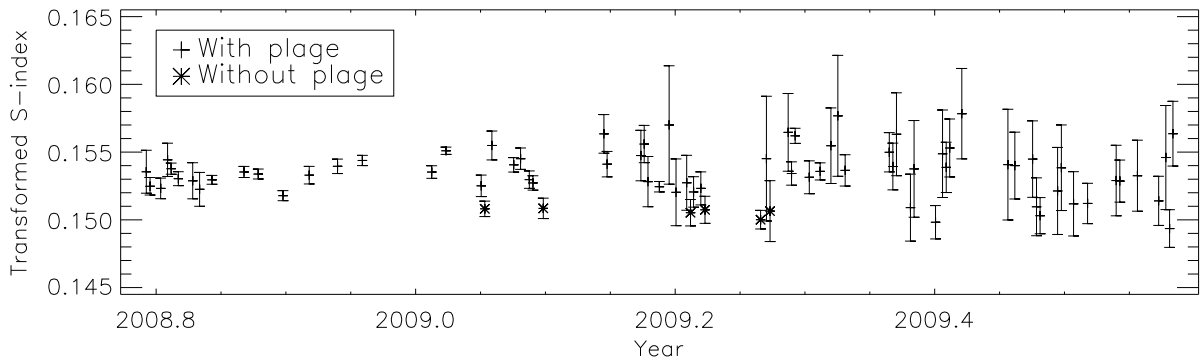}}
\caption{Mt. Wilson S-index measurements obtained for the Sun 
by the Hamburg Robotic Telescope and HEROS \'Echelle spectrograph 
during the extraordinary activity minimum 2008/2009. A minimum
S value of 0.150 is reached on several plage-free days. Error bars 
are mainly due to noise, which varies with temperature and 
humidity.}
\label{band}
\end{figure*}

An important milestone in our understanding of stellar chromospheres
was set by Schrijver (1987) and Schrijver et al. 
(1989, 1992), who identified a so-called basal 
component in the chromospheric line emission, notably in 
the Ca~II~H+K and Mg~II~{\it h}+{\it k} lines, which is relatively 
small compared to and independent of the large observed variations of 
the chromospheric emission commonly attributed to stellar magnetic activity.  
This concept of a basal flux was subsequently confirmed
by several authors, including Strassmeier et al. (1994)
and Pasquini et al. (2000).  A plausible physical
explanation was also readily offered through the dissipation 
of non-magnetic energy,
proposed to exist in the form of acoustic waves.  Hydrodynamic models 
of these waves, including their propagation, shock formation and dissipation,
indeed showed the same dependence of the emitted flux on the stellar 
effective temperature $T_{\rm eff}$ as observed for the basal chromospheric 
flux, which is $T_{\rm eff}^8$ (e.g., Buchholz et al. 1998).
Additionally, in a quantitative study of a distinct Maunder Minimum-type star
in the Mt. Wilson sample, i.e., $\tau$ Ceti, Rammacher \& Cuntz (2003)
were able to reproduce the basal chromospheric flux in terms of
acoustic wave energy dissipation with high accuracy.

Nevertheless, this simple picture was seriously challenged when 
also a minimum flux in the soft X-ray emission was detected, even for 
stars suspected to be Maunder Minimum (``flat activity'') candidates 
of the Mt. Wilson sample (Schmitt 1997; Saar 1998; Judge et al. 2004). 
According to detailed theoretical reasoning
(e.g., St{\c e}pie\'n \& Ulmschneider 1989; Hammer \& Ulmschneider
1991), acoustic waves cannot account for the observed X-ray emission, and 
therefore the exact physical process(es) leading to the basal 
chromospheric flux still remain a topic of intense debate (e.g., 
Judge \& Cuntz 1993; Judge \& Carpenter 1998).

For evolved stars the prevalence of acoustic rather than magnetic heating
appears to be consistent with basic ideas on stellar angular momentum
evolution: when
solar-type stars evolve away from the main-sequence, their rotation
slows down beyond what should be expected from the increase in the 
moment of inertia caused by changes in the internal mass distribution
(i.e., Gray 1991; Schrijver 1993; Schrijver \& Pols 1993; Keppens et al.
1995; Charbonneau et al. 1997).  Hence, the stellar angular
momentum subsides due to magnetic braking resulting from the
onset of massive stellar winds and therefore 
little magnetic activity is expected to persist in evolved stars.

Based on long-term time series observations of chromospheric Ca~II~K emission,
i.e., the Mt. Wilson project (Baliunas et al. 1995), stars identified to
exhibit very low S-index levels, i.e., without regular, cyclic changes,
thus considered as ``flat", appear to be in a state comparable to the
solar Maunder Minimum.  During the period of more than 30 years covered by 
the Mt. Wilson project, which also included several solar activity minima,
the Sun kept a distinct positive distance from these low S-levels; yet is
by wide standards not considered a highly active star.  Therefore,
it remained unclear, how inactive a Maunder Minimum Sun really is.
From historic observations we are aware of a prevailing, though not 
always complete, absence of sunspots, but this may not be representative 
for Ca~II plages.

A rough comparison shows that at activity minimum the solar chromosphere 
is relatively close to its basal flux. This is, however, difficult to verify
for other Maunder Minimum stars, since any quantitative interpretation of
the Mt. Wilson S-indices depends on the exact knowledge of the flux in 
two UV continuum windows used to normalize the chromospheric line emission 
fluxes.  Hence, there may be long-suspected similarities between Maunder 
Minimum 
Mt. Wilson stars, the basal chromospheric flux and the quiet solar chromosphere
(see Judge \& Saar 2007 for measurements in the UV and X-ray regimes), but the 
final proof has still been missing.  In addition, the rich presence of 
mixed-polarity magnetic field in the quiet Sun has always called into question
whether its characteristic physics would be a good proxy for truly inactive 
stars, especially giant stars.

In this paper, we will present the final clue regarding the universality 
of the quiet-Sun chromosphere as the underlying phenomenon to both 
Maunder Minimum-type stars and basal chromospheric flux stars.  We then will
give a quantitative interpretation of the Mt. Wilson S-index data 
in terms of projecting the basal chromospheric Ca~II~K flux into
the S$-(B-V)$ diagram.  Finally, we will discuss the aforementioned 
apparent contradictions in the light of current knowledge 
about the physical processes in the quiet-Sun chromosphere.

\section{The solar S-index at Maunder Minimum level}

For both historic and instrumental reasons, the Mt. Wilson S-index 
is defined as 
the flux ratio between the central 1 {\AA} wide, mostly chromospheric 
Ca~II~H and K line emissions, $F_{\rm H}$ and $F_{\rm K}$, and two nearby 
reference bands of 20 {\AA} of UV quasi-continuum flux, $F_{\rm R}$ and 
$F_{\rm V}$ at 3901 {\AA} and 4001 {\AA}, i.e., 
S = $\alpha\,(F_{\rm H} + F_{\rm K}) /(F_{\rm R} + F_{\rm V})$;
see Baliunas et al. (1995).
According to Duncan et al. (1991), the nightly calibration factor $\alpha$ 
is about constant, i.e., $\alpha = 2.4$.  By contrast, S-indices based on
present, normalized spectra need to be scaled by a factor of approximately 
19 (see Hall et al. 2007, who derived a value 19.2), to compare 
the new S-indices with the original Mt. Wilson 
S-indices, S$_{\rm MWO}$.  In a first test phase of the Hamburg Robotic 
Telescope (HRT) and its \'Echelle spectrograph HEROS (Heidelberg Extended 
Range Optical Spectrograph; see Hempelmann et al. 2005 for a detailed 
description of the HRT including
its hardware) in the time frame between October 2008 and July 2009
we obtained normalized spectra of a sample of stars and the Sun spanning
the whole activity scale 
with a spectral resolution of $R \approx 20,000$.
Using specifically observations of 29 selected, well-observed and stable
Mt. Wilson stars with S$_{\rm MWO}$ between 0.15 and 0.7, 
we determined the following, very tight linear relation between ``classic'' 
S$_{\rm MWO}$-indices and our own, raw S$_{\rm HRT}$-indices via
${\rm S}_{\rm MWO}~=~(0.0294 \pm 0.0040) + (19.79 \pm 0.56) \cdot {\rm S}_{\rm HRT}$,
which allows us to compare our S-indices to the large number of 
MWO S-indices (Duncan et al. 1991).

During October 2008 and July 2009, we observed the Sun on 73 occasions, usually
by taking day-sky spectra and twice by taking lunar spectra. 
Fortuitously, this observing series coincides 
with the exceptionally long and deep past solar minimum. All HEROS spectra 
were reduced with the fully automatic reduction pipeline of the HRT 
(see Mittag et al. 2010). The day-sky spectra were furthermore corrected 
for Mie and Rayleigh scattering, with a lunar spectrum serving as reference. 
To estimate the error of this correction, the continuum flux of the 
day-sky spectra, after correction, and the reference moon spectrum 
were compared. The remaining differences are very small, indeed, in the 
range of $0.2\%-1.8\%$, on average approximately 0.6\%. 

Our individual solar measurements are shown as a time series in Fig. 1; 
the measured S-index values are distributed between 0.150 and 
0.157, with an average solar S-index of 
${\rm S}_{\rm{MWO}} = 0.153 \pm 0.005$ in that period.
By contrast, the solar S-index measurements
obtained in the context of the Mt. Wilson project showed that
the solar S-index varied between about 0.16 (normal activity minima) 
and 0.22 (maxima) between 1966 and 1993 (Baliunas et al. 1995). 
The abnormally low value of the solar ${\rm S}_{\rm{MWO}}$ 
found by us in 2009 must be real and caused by the unusually deep solar minimum, 
rather than by errors in the above-stated transformation from 
${\rm S}_{\rm HRT}$ to ${\rm S}_{\rm MWO}$, since 
the observed, average offset ($>0.007$) is a lot larger than the 
respective uncertainty of the transformation ($0.004$).

In summary, we argue that our solar S-index measurements from
October 2008 to July 2009 are indeed lower than any of the solar
S-indices measured in the context of the Mt. Wilson project, ever.
Most interestingly, these values are very close to the S-indices of some 
solar-type, ``flat activity'' Mt. Wilson sample stars, i.e., 
HD 43587 ($B-V=0.61$) and HD 143761 ($B-V=0.60$; see Baliunas et al. 1995).  
The absence of any higher activity for over 30 years make these stars 
excellent Maunder Minimum candidate stars, and furthermore, because their 
$T_{\rm eff}$ values are very similar to those of the Sun, their S-indices can 
be directly compared to the solar values.

Since sunspots were very rare during our specific observing period, there
is little correlation between the S-index values and the sunspot numbers.  
In particular, there are many S-index measurements on days where the 
sunspot number was zero.
However, for the solar Ca~II emission the sunspots are of lesser importance,
whereas the presence of any plages (faculae) would directly impact the
measured S-index. Indeed, the lowest obtained solar S-values of nearly 0.150
occurred only on days without {\it any} plages at all (see Fig.~2), as 
already proposed by White et al. (1992). 

\begin{figure}  
\centering
\resizebox{0.8\hsize}{!}{\includegraphics{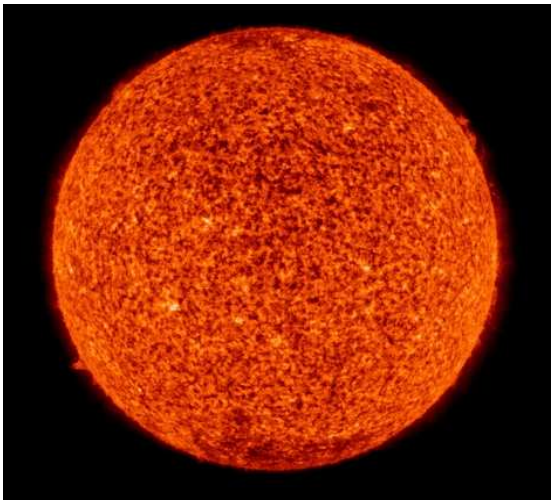}}
\caption{Plage-free solar chromosphere on February 6, 2009, 
taken by SOHO at 304 nm.}
\label{band}
\end{figure}

\section{S-index and basal chromospheric flux}

Next we explore the relationship between the S-index and the basal chromospheric
flux for main-sequence stars and giants.  Figure 3, which displays the distribution
of S$_{\rm MWO}$ (Duncan et al. 1991) for main-sequence stars (black dots) and
giant stars (shaded dots) over $B-V$, strongly suggests that the lower boundaries
of the respective distributions correspond to inactive stars with pure basal
chromospheric emission.  Main-sequence stars of spectral types K and M toward
the red end are increasingly more active.  By contrast, only G-type giants
show a significant amount of chromospheric activity (e.g., 
P{\'e}rez Mart{\'i}nez et al. 2011). 
It is also at the basal flux edge (around S=0.15) of G-type 
main-sequence stars, where the Maunder Minimum-type ``flat activity'' 
Mt. Wilson stars as well as the plage-free Sun are found.  However,
because quasi-continua and Ca~II photospheric line-cores are both included
in the definition of S and also depend on the stellar gravity, giants at
basal flux levels are offset from the respective main-sequence to slightly
lower S-values.

With a detailed revision of the Ca~II~K basal chromospheric surface line 
flux, based on new spectroscopic observations and non-LTE photospheric models, 
which are still being developed, we employ a simple relationship based on 
Mg~II~{\it h}+{\it k} line emission, which is much easier to measure and 
quantify. Using IUE spectra of a set of inactive cool giant stars, 
we previously derived
the following relation between the ``basal'' chromospheric surface 
flux in the Mg~II {\it h}+{\it k} lines and effective temperature $T_{\rm eff}$:
$\log{F_{\rm Mg~II}}~=~7.33\log{T_{\rm eff}} - 21.75$ (in cgs units;
see P{\'e}rez Mart{\'i}nez et al. 2011 and references therein, for details).
Earlier work on main-sequence stars arrived at a very similar result, which
again shows that the basal chromospheric flux itself (within its 
uncertainties) does not depend on gravity.
Furthermore, according to Schrijver et al. (1992), Mg~II~{\it h}+{\it k} and Ca~II~H+K 
chromospheric line fluxes differ by a fixed factor, which is $0.78 \pm 0.05$. 
By dividing the Mg~II fluxes from the above given relationship by this factor,
we are able to derive Ca~II~H+K basal surface fluxes, allowing us to compute
``basal'' S-indices. 

Compared to the work by 
Strassmeier et al. (1994), who determined $\log{F_{\rm Ca~II}} = 
8.0\log{T_{\rm eff}} - 24.8$ directly from Ca~II line emissions of 
cool giant stars, our basal flux values turn out to be higher by a factor 
of 2 to 4; they are of the same order as the photospheric line core fluxes
as confirmed by Figure 4. This plot compares the Ca~II~K line core of 
HD~109379, a basal flux giant, and its chromospheric emission at basal 
flux level to a synthetic spectrum given by a stellar atmosphere
model obtained by the Phoenix code (see Hauschildt \& Baron 2005; with
$T_{\rm eff} = 5800$~K and $\log{g} = 2.5$) to make a photospheric template 
and an accurate, physical surface flux scale.  By contrast, the 
scale used by Strassmeier et al. (1994) relied on historic work 
from the 1970s.  The same appears to apply to the work by White et al. (1992),
who found that the Ca~II~K line flux of inactive solar-type chromospheres would 
fall much below that of the average quiet Sun.

\begin{figure}  
\centering
\resizebox{0.9\hsize}{!}{\includegraphics{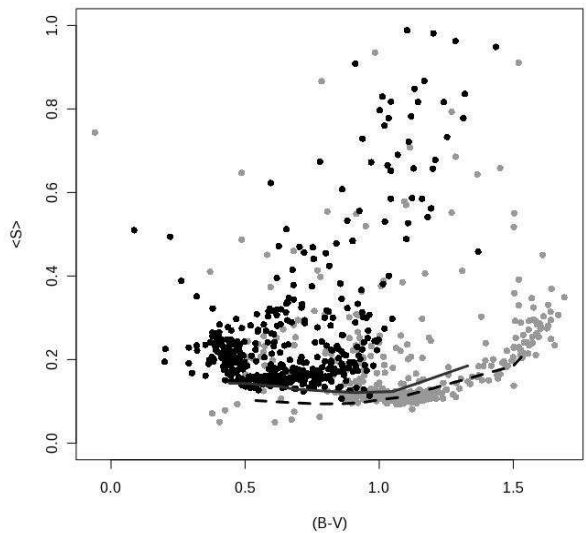}}
\caption{Observed distribution of stellar Mt. Wilson S-indices over $B-V$,
together with the location of basal chromospheric Ca~II flux (see text). 
Although exhibiting the same basal flux, giants (shaded dots) show
a systematic offset in S against main-sequence stars (black dots) because
their photospheric spectral properties are gravity-sensitive
(dashed and solid line, respectively).  Most notably,
Mt. Wilson Maunder Minimum-type stars and the plage-free solar S-values
coincide with the Ca~II basal flux limit.}
\label{band}
\end{figure}

\begin{figure}  
\centering
\resizebox{0.9\hsize}{!}{\includegraphics{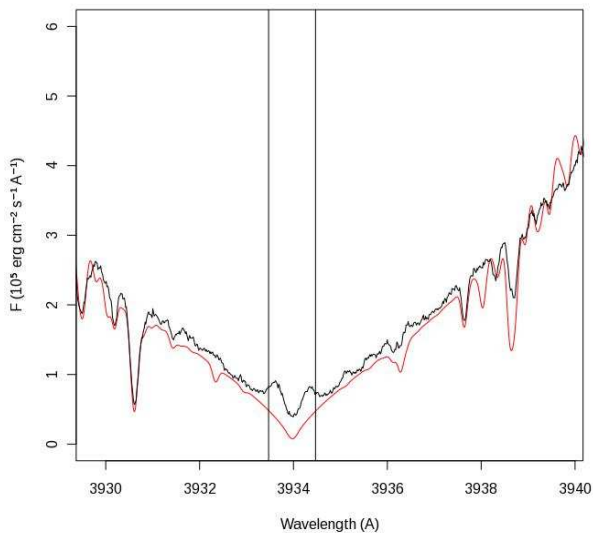}}
\caption{Ca~II~K line core of HD 109379, a giant star, showing its basal
chromospheric line emission in nearly equal proportion to the underlying 
photospheric line flux, represented by a Phoenix model atmosphere spectrum
(Hauschildt \& Baron 2005).}
\label{band}
\end{figure}

To properly translate the basal chromospheric Ca~II~K flux
into S$_{\rm MWO}$, the underlying photospheric fluxes 
in the Ca~II line centers, $F_{\rm H}$ and $F_{\rm K}$, need to be known, 
as well as the above-mentioned UV quasi-continuum fluxes $F_{\rm R}$ and 
$F_{\rm V}$. For a variety of effective temperatures and two representative 
sets of gravity (i.e., $\log{g} = 4$ for the old, least active
main-sequence stars 
and $\log{g} = 2$ for LC~III giants) we derived these quantities from 
Phoenix model spectra. The resulting raw values of ``basal S'' were multiplied 
by 19 to match the S$_{\rm MWO}$-scale.  Finally, effective temperatures were 
translated into $B-V$ colors using the tables of Schmidt-Kaler (1982).
The resulting ``basal'' S-values are depicted in Fig. 3 (solid and dashed
lines, respectively). 
Since both quasi-continua and Ca~II photospheric line cores, 
are gravity-sensitive, giants at the basal flux limit are offset
from the main-sequence stars at the basal flux limit to slightly 
lower S-values.
It is near the main-sequence basal flux S-values (around 0.15) 
that we find the Maunder Minimum type ``flat activity'' stars
as well as the plage-free Sun.

\section{A local ``fast'' dynamo: new answers to old questions?}

Through the kind cooperation of our central star in 2009, we found  
direct proof that it is the quiet-Sun chromosphere which is behind
the formation of basal chromospheric flux, and that a Maunder Minimum-type 
star exhibits a truly quiet-Sun chromosphere.  Therefore, we may now conclude that
the quiet Sun is not just similar to inactive stars, but it even 
represents the universal phenomenon that produces the basal flux of all 
inactive cool stars with some adjustments for low gravity pertaining to 
giant stars (see discussion). With this insight in mind, can we now learn 
something new by looking at the physical processes typical for the 
quit-Sun chromosphere?

The rich presence of mixed-polarity flux in the quiet Sun nurtures
phenomena like spicules and bright points, which are related to rising flux-tubes 
and their dynamic interaction with convection, and a minimal coronal 
emission of soft X-rays from solar coronal holes and even Maunder Minimum 
stars (Schmitt 1997). It was long thought that this magnetic presence was 
a side product of the main stellar activity through magnetic dissipation 
of the strong, rising local fields. But the level of magnetic
flux-tube density in the quiet-Sun surface does {\it not} change with the
overall activity of the Sun, while we would expect it to show some 
following-tendency with the solar activity on a time-scale of months
or years. Hence, its nature must be entirely independent of what 
we understand as magnetic activity (local, strong fields, rising up from 
the bottom of the convection zone where a global dynamo produced them).

Recently, V\"ogler \& Sch\"ussler (2007) have presented an MHD simulation
for the quiet-Sun surface, which yields a mixed-polarity field, generated 
by a local ``fast'' turbulent dynamo operating only in the thin uppermost 
layer of the convection zone. The principle of such a non-cyclic mechanism 
to create a light distribution of random magnetic field has already been 
suggested by Saar (1998), who also anticipated its potential in explaining
the soft X-ray flux and transition region emission concerning inactive 
chromospheres,
where pure acoustic wave heating would well account for only the basal line 
flux. The simulation of V\"ogler \& Sch\"ussler (2007) starts with a weak 
seed field, and soon a fairly well-defined saturation level of the magnetic 
energy density of a few percent of the total kinetic energy is attained. 

This result is nicely consistent with the relative flux contribution of the 
solar bright points to the total light, in both MHD simulations (Unruh et al. 
2009) and high-resolution 
observations (by the SUNRISE balloon-borne observatory, Riethm\"uller et al. 
2010). In Ca~II~K, bright points appear to contribute only about  
10\% of the quiet-Sun chromospheric flux. The by far largest contribution 
to the quiet-Sun chromospheric line emission seems to arise from a highly 
dynamic process (W\"oger et al. 2006), which operates on a time-scale 
significantly shorter than the underlying convection, and which we may 
therefore attribute to the action of acoustic waves; see, e.g., Cuntz et al.
(2007) for recent solar acoustic heating models.

Hence, both present quiet-Sun observations and detailed hydrodynamic 
case studies of non-active chromospheres with acoustic wave dissipation 
strongly hint at kinetic energy to be responsible for creating the 
major part of the basal chromospheric flux.   
However, this concept alone cannot explain
the coronal soft X-ray emission from above non-active areas. 
However, invoking the idea of an embedded fairly 
constant share of magnetic field, this puzzle is finally solved: 
the typical minimum soft X-ray surface flux, as found in 
coronal holes and for Maunder Minimum-type stars like $\tau$ Cet, 
is close to 10$^4$ erg~cm$^{-2}$~s$^{-1}$ (Schmitt 1997; Judge et al. 
2004). That is nearly two orders of magnitude below the total 
basal chromospheric
cooling flux of all emission lines combined, consistent with an energy 
content of the magnetic field of a few percent of the kinetic energy of 
the turbulent chromosphere. According to Pevtsov et al. (2003), the minimum 
X-ray flux requires a magnetic field of about only 10 Gauss, which 
easily falls within the yield of the MHD simulation of V{\"o}gler \& 
Sch{\"u}ssler (2007). 

We therefore conclude that local dynamo action finally resolves the 
long-standing puzzle of the origin of a minimum X-ray flux, in the 
absence of any (``classical'') stellar activity as we used to know 
it, i.e., in the absence of active regions. 

\section{Discussion and conclusions}

In view of the universality of physics, it is a sensible speculation that 
the same local ``fast'' dynamo would produce a similar content of 
mixed-polarity magnetic flux-tubes in giant chromospheres. Since the 
local ``fast'' dynamo is entirely different from the classic 
$\alpha - \Omega$ dynamo 
and therefore independent of any global dynamo action, it should 
provide a universal presence of flux-tubes at a fairly well-defined 
saturation level, for every cool giant, and not just for their highly 
active tail-end (e.g., Konstantinova-Antova et al. 2010). Indeed,
according to Brandenburg (2011), these locally generated magnetic fields
do not vary much with Prandtl number.

There is one special aspect for giant stars, however, which requires more careful 
consideration: Here we must expect all scale lengths $l$ to be a lot larger 
because of the much lower gravity, following Schwarzschild's simple 
relation $l \propto R/g$ with $R$ and $g$ as stellar radius and
surface gravity, respectively (Schwarzschild 1975). In this concept   
of a growing extent of chromospheres and their embedded features with lower 
values of $g$, 
we would also expect the occurrence of giant spicules.  Indeed, these have
already been suspected two decades ago based on data gathered during 
spectroscopic eclipses of HR~2554 and HR~6902 (Schr\"oder \& H\"unsch 1992).

Furthermore, if the local ``fast'' turbulent dynamo were still able to operate
in an environment of extremely low gravity, particularly concerning the
photospheres of K-type supergiants like 32~Cyg, the resulting presence of 
local, random magnetic fields would also provide us a plausible explanation 
for transient, prominence-like clouds of chromospheric material, 
long known from spectroscopic eclipse observations of $\zeta$~Aurigae
systems by ``satellite lines'' in Ca~II~K (e.g., Wilson 1960). 
Schr\"oder (1983) found a large cloud in an IUE spectra series of the 
1981 eclipse of 32 Cyg, 1/6 giant radius off-limb, by its Rayleigh 
scattering and concluded that a magnetic field of about 4 Gauss 
(if $T_{\rm eff} = 3000$~K) would suffice to balance its thermal energy. 
With an updated choice of a chromospheric temperature of 
6000~K, this would amount to about 6 Gauss, which still lies very reasonably 
within the the means of the quoted local dynamo.

A supply of flux-tubes to giant chromospheres would also provide a 
natural explanation for the vertical transport of mechanic flux 
from the kinetic energy reservoir of the turbulent chromosphere up into 
the cool wind acceleration region, as suggested by Schr\"oder \& Cuntz 
(2005) and Suzuki (2007).  The problem is that acoustic waves are
expected to be insufficient because they dissipate their energy relatively
close to the star (e.g., Cuntz 1990), but vertical magnetic structure
may constitute a venue for this kind of energy transport.
Hence, we conclude that a number of hitherto unresolved problems
can be seen in an entirely new light, and apparent contradictions
disappear if the latest insight into the principal and apparently
quite universal physics of the quiet-Sun chromosphere is considered. 

Another very interesting and direct result of our 2009 solar S-index 
measurements is the correspondence of the established Maunder Minimum 
S-level of 0.150 with the complete absence of even plages, hence of any sign 
of activity regions. This implies that during a Maunder Minimum, surface
activity completely vanishes for most of the time or must be very subtle. 
The question remains as to why and for how long.

\begin{acknowledgements}
This work was supported by travel funds from the
bilateral Conacyt-DFG grant No. 147902.  Furthermore, we are most
grateful for the funding of the robotic telescope project by the 
University of Hamburg and the Deutsche Forschungsgemeinschaft 
(instrumentation), by the Universities of Guanajuato and Li{\`e}ge,
and by the State of Guanajuato (infrastructure at the new location in 
Guanajuato, Mexico).  Finally, we wish to thank our colleague Peter 
Hauschildt for providing us with a large set of Phoenix models.
\end{acknowledgements}

\end{document}